\documentclass[prl,aps,floatfix,a4paper,reprint]{revtex4-1}

\usepackage{ulem}
\usepackage{graphicx}
\usepackage{amsmath}
\usepackage{float}
\usepackage{placeins}
\usepackage{bm}
\usepackage{xcolor}
\begin{document}

\title{Self-Induced Rayleigh-Taylor Instability in Segregating Dry Granular Flows}
\author{Umberto D'Ortona}
\email{umberto.d-ortona@univ-amu.fr}
\affiliation{Aix Marseille Univ., CNRS, Centrale Marseille, M2P2, Marseille, France}
\author{Nathalie Thomas}
\affiliation{Aix Marseille Univ., CNRS, IUSTI, Marseille, France}
\date{\today}
\begin{abstract}

Dry granular material {\color{blue} flowing on rough inclines can experience
a self-induced Rayleigh-Taylor (RT) instability followed by the spontaneous
emergence of convection cells. For this to happen,}
particles are different in size and density,
the larger particles are the denser but still segregate toward the surface. 
When the flow is, 
{\color{blue} as usual}, initially made of two
layers, dense particles above, a Rayleigh-Taylor instability develops during
the flow. {\color{blue} When the flow is initially made of one homogeneous 
layer mixture,} the granular
segregation leads to the formation of an unstable layer of large-dense particles at the
surface which subsequently destabilizes {\color{blue} in a RT plume pattern.  The unstable
density gradient has been only induced by the motion of the granular matter.}
This self-induced Rayleigh-Taylor
instability and the two-layer RT instability are studied using two different methods, experiments and
simulations.  At last, contrarily to the usual fluid behavior where the RT
instability relaxes into two superimposed stable layers of fluid, the granular
flow evolves to a pattern of alternated bands corresponing to recirculation cells
analogous to Rayleigh-B\'enard convection cells {\color{blue}where segregation 
sustains the convective motion.}

\end{abstract}

\maketitle

The Rayleigh-Taylor instability is one of the most commonly studied
hydrodynamical instability. It occurs when a dense fluid is put atop a lighter
fluid \cite{Chandrasekar,Charu}. This phenomenon is encountered in various
fields like volcanoes, supernovae explosions or while pouring vinegar over oil
in your kitchen. Another frequently studied instability is the
Rayleigh-B\'enard instability. It occurs when a horizontal layer of fluid
is heated from below \cite{Getling}. 
{\color{blue} In both cases, the reason for instability
or convection is external: the two layers have been superimposed, or  heat
is brought into the system. But in this letter, the granular flow spontaneously
creates its unstable state, then the instability happens,  and the flow sustains the convective state without any
external cause.}

Dry granular material behaves as liquid when put into motion
\cite{Duran,Herrmann,Andreotti}.
One of the most striking phenomenon is the granular segregation
(Brazil Nut effect): when particles of different
sizes flow together, large particles migrate to the flow surface \cite{Jaeger,Thomas00,Shinbrot,ThomasDOrtona18}.
{\color{blue} This process results from a grain-scale interaction between 
large and small particles {\color{blue} during the granular flow}. Moreover, it vanishes for a large (resp. small) 
particle surrounded only by large (resp. small) particles.}
Another segregation occurs when particles having
different densities flow together: denser particles migrate to
the bottom \cite{OttinoKhakhar00,Thomas04,OttinoLueptow,TripathiKhakhar13}. Depending
on size and density ratios, large-dense particles could sink or raise. Here,
they are chosen such that large particles raise. {\color{blue} With this choice,  
segregation pushes the denser particles toward the surface and 
creates a reverse, and unstable, density gradient.} {\color{blue}The system induces its own unstable state simply by flowing, and 
not because of external causes. The unstable state is self-induced by the flow 
which is unusual in Fluid Mechanics.}

To our knowledge, the Rayleigh-Taylor instability between two dry 
granular materials of different densities has never been studied, even though
several works report the study of a Rayleigh-Taylor instability involving
a granular material and a fluid, liquid or air
\cite{VoltzPesch01,CarpenBrady02,VinninglandFlekkoy07}. 
This letter first shows the existence of an instability between two 
initially superimposed dry
granular layers flowing down an incline. Buoyancy acts there as it does in liquids.
Second, and more interestingly, it shows
that when particles differ in size, a self-induced Rayleigh-Taylor instability
may arise: {\color{blue} one initially homogeneous mixture layer spontaneously develops an unstable state when flowing. The segregation leads to the}
formation of a layer of
large-dense particles at the surface that will subsequently destabilizes through a RT instability. 
This new phenomenon involves both {\color{blue} an individual behavior of particles through segregation and a collective behavior
through a hydrodynamical destabilization of the dense upper layer}.

Moreover, when the granular media goes on flowing, a third 
very surprising phenomenon occurs at long time: the granular
flow evolves to a pattern of alternated bands with recirculation cells
analogous to Rayleigh-B\'enard convection cells.
Convection has been observed in rapid granular flows for which the granular 
temperature plays the role of the temperature in a liquid \cite{ForterrePouliquen01,BorzsonyiEcke09}, but here, a moderate slope is used and the 
flow is dense. The motor of the convection is not the 
temperature, but the segregation which is induced by the flow itself.
{\color{blue} The convection is then self-sustained by the flow because,
during the flow, segregation and buoyancy compete. 
Our system has similarities with bioconvection induced by upwardly self-propelled denser micro-organisms \cite{HillPedley05}, except that our particles are neighbors-propelled instead of self-propelled.}


As the three phenomena reported here have never been studied or observed, 
two methods of investigation were necessary.  Experiments (later in the text) and simulations
have been conducted. 
Numerical simulations are performed using the Distinct Element Method (DEM).
The normal force is modeled using a damped linear spring. The tangential
force is of Cundal and Strack type \cite{Dippel,DOrtonaThomas16}.
Two types of particles are used. The properties of the small particles
are those of cellulose acetate:
density $\rho=1308$~kg~m$^{-3}$, restitution coefficient $e=0.87$,
friction coefficient $\mu=0.7$, and a diameter of $d=6$~mm.
Large particles have the same friction and restitution coefficients,
but size $d_l$ and density $\rho_l$ are adjusted depending on the needs. To prevent
crystallization, each species present a uniform size distribution ranging
from 0.95$d$ to 1.05$d$. The collision time is $\Delta t$ =10$^{-4}$~s, consistent with previous simulations and sufficient for modeling hard
spheres \cite{DOrtonaThomas16,Ristow00,Campbell02,SilbertGrest07}.
{\color{blue} These parameters correspond to a
stiffness coefficient $k_n = 7.32\, 10^4$~N~m$^{-1}$ and a damping
coefficient $\gamma_n = 0.206$~kg~s$^{-1}$ \cite{Dippel,DOrtonaThomas16}. The integration time step
was $\Delta t/50 = 2\,10^{-6}$~s to meet the requirement of numerical
stability \cite{Ristow00}.}
Rough inclines are modeled using a monolayer of bonded small particles
placed randomly leading to a compacity about 0.57. These particles
have an infinite mass and do not move during the simulation. Flowing
particles are randomly placed on the rough incline, either on a
two-layer (large particles above) or in one homogeneous mixture layer 
configuration.
At time zero, gravity is set with a tilt angle of  $\theta=23^\circ$ and
the flow starts. Periodic boundary conditions are applied in the two directions
parallel to the incline ($x-y$). 


\begin{figure}
\includegraphics[width=0.99\linewidth]{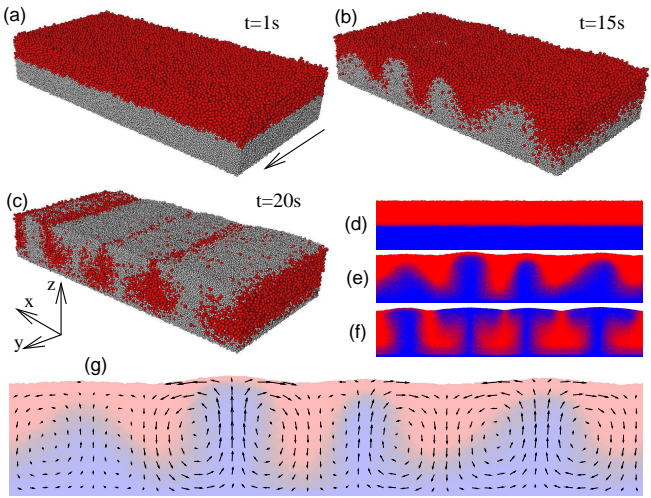}
\caption{Rayleigh-Taylor instability in a flow initially organized in two layers.
(a-c) Successive pictures of the destabilization (large particles in red). 
The arrow indicates the direction of the flow. 
(d-f) Concentration fields ($x-z$) averaged over the flowing direction $y$ corresponding to figures (a-c), 
{\color{blue}(g) transverse velocity field, in the plane $x-z$, indicated
by arrows corresponding to figures (b) and (e).}}
\label{huge36h15dsimuprofil}
\end{figure}
The case of an initial two-layer system, with large-dense 
particles forming the upper layer, is first studied. 
Figures~\ref{huge36h15dsimuprofil}(a-c) present
successive pictures of the numerical simulation. 
The thickness of the flow is $H=36d$,
the length in the flowing direction is $L=100d$, and the width is $W=200d$. 
Flowing particles have a size ratio $d_l/d=2$, 
a density ratio $\rho_l/\rho=1.5$ and an equal volume
fraction.
After the granular material has started to flow, the interface between 
the two species destabilizes ($t=15$s) and forms a plume pattern ($t=20$s)
{\color{blue} typical of a Rayleigh-Taylor instability obtained with viscous liquids having
a viscosity ratio close to 1 \cite{woidt78}} (See Video 1 in Supplemental Material
at [URL will be inserted by publisher]). As the flow stretches the interface in the $y$ direction, the plumes take the shape of parallel stripes \cite{YuChang97}. The plume pattern is clearly
visible in vertical concentration
fields obtained by averaging particle volume
fraction in the flow direction (Fig.~\ref{huge36h15dsimuprofil}(d-f)).  
{\color{blue} Plumes  have rising/descending motion through the
whole thickness and spreading heads at top and bottom boundaries. The transverse
velocity field (Fig.~\ref{huge36h15dsimuprofil}(g)) is typical of a RT instability with vertical flow inside plumes and
contra-rotative rolls between plumes.} A Rayleigh-Taylor
instability between two granular materials occurs and for these chosen size and
density ratios ($d_l/d=2$ and $\rho_l/\rho=1.5$), the granular segregation does not prevent it.
 The width of the flow being $W=200d$, the wavelength can be estimated
$\lambda\simeq 1.4H$. An other simulation with $H=20d$ has given a wavelength
of $\lambda\simeq 1.7H$.
At the end of the simulation ($t=70$s), contrarily to the usual RT instability
in liquids where the system relaxes into two superimposed layers at rest,
the granular flow reaches a pattern of parallel stripes made of pure
large particles alternating with stripes made of a mixture of small and
\begin{figure} 
\includegraphics[width=0.99\linewidth]{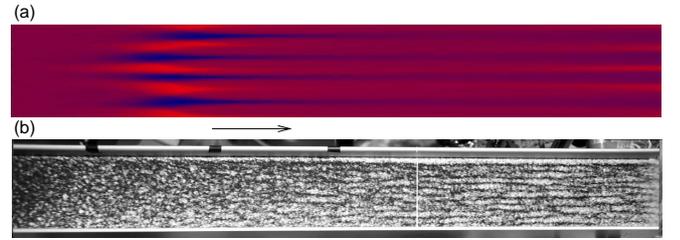}
\caption{Rayleigh-Taylor instability in a flow initially organized in two
layers.  (a) Space-time diagram viewed from the top, the concentration is
obtained by averaging over the thickness of the flow, time passes from left to
right up to 70~s. (b) Top view of the corresponding experiment. The flow is
from left to right (see arrow), large particles are black. Successive stages of
the destabilization are visible from left to right.} 
\label{huge36h15dspatio}
\end{figure}
large particles. This pattern is visible in
Fig.~\ref{huge36h15dspatio}(a) where a space-time diagram of the instability is
made.
		
Experiments have been conducted on a 110~cm long and 6.85~cm wide rough incline
(Fig.~\ref{huge36h15dspatio}(b)).  Flowing particles are ceramic beads: white
Zirshot (diameter $d=250-280\mu$m and density $\rho=3850$~kg~m$^{-3}$) and
black Cerabeads (diameter $d_l=500-560\mu$m and density
$\rho_l=6200$~kg~m$^{-3}$) inducing a size ratio of $d_l/d=2$ and a density
ratio of $\rho_l/\rho=1.61$.
{\color{blue} With the chosen thicknesses, the aspect ratio $W/H$
is slightly larger in the experimental case than in numerical simulations.  The
length of the channel corresponds to the duration time in simulations
with periodic boundaries and is smaller. It is difficult to reach long time
evolutions in the experiments.} The inclined plane is first placed horizontally
and covered with two superimposed layers of small-light (at bottom) and
large-dense particles.  The incline is then slowly tilted at 23$^\circ$, the
gate at the bottom end of the incline serves as a containment.  At $t$=0, the
gate is removed, and the flow triggering rapidly spreads up the slope. The flow
starts everywhere with a small time delay (See Video 2 in Supplemental Material
at [URL will be inserted by publisher]).  Figure~\ref{huge36h15dspatio}(b) is taken when the
triggering reaches the upper part of the inclined plane at the left end.  All
successive stages of the Rayleigh-Taylor instability can be seen from left to
right: two layers where only the upper black layer is visible, white dots
showing ascending small-white particle plumes, and organization of the system
toward a band pattern.  Several experiments has been performed with a flow
thickness, measured with the deflection of a laser sheet, from $H=12\,d$ to
$19\,d$.  The wavelength of the band pattern is found in between 1.67$H$ and
1.93$H$.

\begin{figure}
\center
\includegraphics[width=0.85\linewidth]{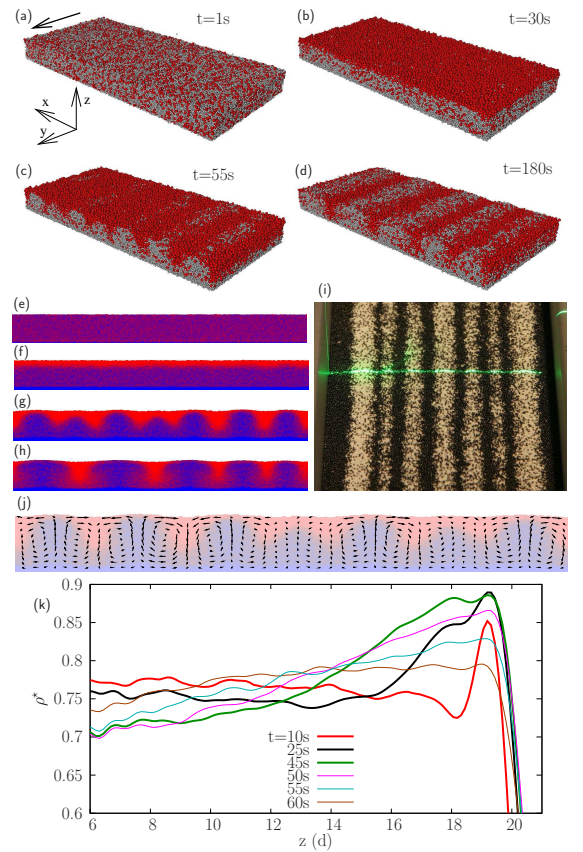}
\caption{Self-induced Rayleigh-Taylor instability starting from 
one initially homogeneous layer.
(a-d) Successive pictures of the destabilization (large particles in red).
(e-h) Concentration fields ($x-z$) averaged
in the flowing direction $y$ corresponding to (a-d) 
(i) Picture of the lower part of the experiment. Small variations of flow 
thickness showed by
the laser sheet deflection: dark bands made of large particles are thinner.
{\color{blue}
(j) Transverse velocity field indicated
by arrows, corresponding to figures (c, g).
(k) Successive reduced density profiles near the surface (at z=20$d$). Instability starts between 45 and 50 s.}}
\label{50p20hr15dHuge}
\end{figure}

%
The case of an initially-homogeneous one-layer system is now considered {\color{blue} numerically and experimentally}.
Figures~\ref{50p20hr15dHuge}(a-d) show the time evolution of the numerical 
simulation of a mixture of large-dense and small-light particles flowing on an 
incline. The
thickness of the flow is $H=20d$ to reduce computing time, other physical 
parameters are kept
identical to Fig.~\ref{huge36h15dsimuprofil}. In between $t=0$ and $t\simeq 40$s, the segregation 
induces the formation of a layer of large-dense particles at 
the surface.  Then ($40$~s~$\lesssim t\lesssim 70$~s), the surface layer 
destabilizes and organizes in stripes parallel to the flow (see Video 3 in Supplemental Material at [URL will be inserted by publisher]).
The vertical concentration fields (Figs.~\ref{50p20hr15dHuge}(e-h)) show
the formation of the layer of large-dense particles, its destabilization
and the formation of a plume pattern typical of a Rayleigh-Taylor instability.
{\color{blue} The transverse velocity field after destabilization is similar to
the one of the initially two-layer system. It presents vertical flows aligned
with plumes typical of a RT instability and contra-rotative rolls 
between plumes (Fig.~\ref{50p20hr15dHuge}(j)). 
Plumes go through the whole thickness and rolls take
place on the whole thickness as well. Rayleigh-Taylor instabilities are
associated with an unstable density gradient.
Figure \ref{50p20hr15dHuge}(k) shows vertical profiles of the reduced density
$\rho^\star$ which is the bulk density divided
by the density of small-light particles. 
In our simulations, we measured for a region full of small particles
a reduced density $\rho^\star=0.59$
(corresponding to the volume fraction of a random loose packing \cite{Dullien92})
for large-dense particules $\rho^\star= 0.88$, and for a 50\% mixture $\rho^\star= 0.78$.
Successive density profiles show the accumulation of 1 to 2 pure layer of large beads at the
surface, an increased concentration of these beads near the surface, and the
formation of an unstable density gradient (Fig. \ref{50p20hr15dHuge}(k)).
The maximal gradient takes place from a value of $\rho^\star= 0.88$ (large beads) to
0.75 (mixed system) on around $8d$ thick.} {\color{blue} When destabilization starts (between
45 and 50 s), the gradient rapidly vanishes when plumes cross over.}
The steady regime (Figs.~\ref{50p20hr15dHuge}(d,h)) will be discussed
at the end of this letter. Looking carefully at the free surface of the flow
(Fig.~\ref{50p20hr15dHuge}(h)), one can see that the surface is not 
flat, stripes of large particles corresponds to hollows, and stripes of
the mixture of small and large particles are bumpy. The same
phenomena can be seen experimentally using the deflection of a laser sheet 
(Fig.~\ref{50p20hr15dHuge}(i)). Dark bands of large particles correspond to 
depressions {\color{blue} around 0.1~mm deep.}

\begin{figure}
\center
\includegraphics[width=0.99\linewidth]{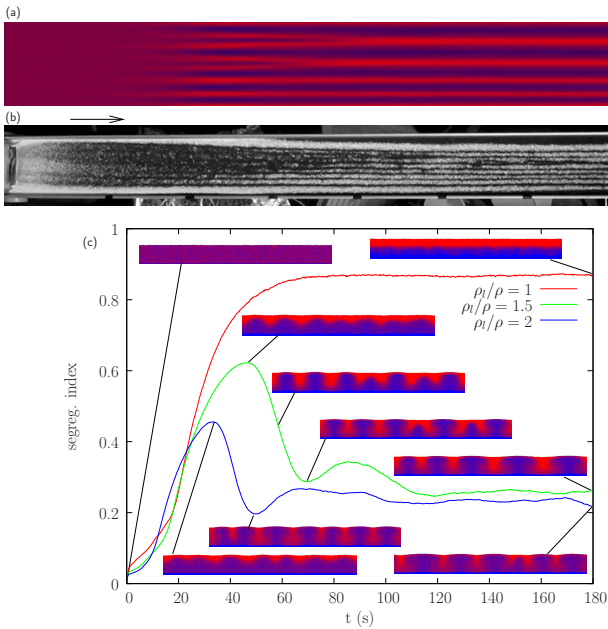}
\caption{Self-induced Rayleigh-Taylor instability of an initially homogeneous layer.
(a) Space-time diagram of the simulation ($\rho_l/\rho=1.5$) averaged over the thickness of the flow. Time passes horizontally up to 180~s. 
(b) Picture of the whole experiment ($\rho_l/\rho=1.61$): a homogeneous mixture is injected at 
the top (left) of the incline, then segregation appears, and the instability induces the formation of a band pattern.
(c) {\color{blue}Time evolution of the segregation index and 
corresponding vertical concentration fields illustrating the evolution of the 
instability. Two other density ratios ($\rho_l/\rho=1$ and 2) are added for comparison.
With no density difference ($\rho_l/\rho=1$) no instability occurs}.  
}
\label{50p20hr15dspatio}
\end{figure}

Experimentally, for
the initially homogeneous flow (Fig.~\ref{50p20hr15dspatio}(b)),
the incline is initially empty and a feeding
container is added at the left end (not seen). {\color{blue} The incline is reduced
to 91~cm due to the container} which
is filled with about 400 alternated thin layers of small and large
particles to provide a feeding as homogeneous as possible.
At $t=0$ the lock gate of the container is opened and the flow starts.
The picture is taken when a stationary regime is reached.
Along the incline, all phases of the instability are visible:
the granular flow is initially homogeneous
($0\lesssim x\lesssim 10$~cm), granular segregation drives large particles
to the surface ($10$~cm$\lesssim x\lesssim 20$~cm), destabilization occurs
($20$~cm $\lesssim x\lesssim 40$~cm) and the flow organizes in a pattern of parallel
stripes ($x\gtrsim 40$~cm) (see Video 4 in Supplemental Material at [URL will be inserted by publisher]). For a quantitative comparison,
the length of the experiment $L\simeq 3400d$ corresponds
in the simulation to the flowing time $t\simeq $~60s, thus to one third 
of the space-time diagram (Fig.~\ref{50p20hr15dspatio}(a)).
Several experiments have been performed and give a wavelength for the
initial destabilization that ranges from 
$\lambda\simeq 1.75H$ to 1.93$H$. The measured flow thicknesses are from $H=9d$ 
to 16$d$.
The spreading of the experimental measurements is
due to thickness irregularities.
Nevertheless, that is in good agreement with the wavelength obtained
in the simulation $\lambda\simeq 40d\simeq 2\,H$.


In granular flows, the evolution of the segregation is quantified using a 
segregation index $SI$:
$$ SI=2\frac{CM_l-CM}{H}, $$
with $CM_l$ and $CM$ the vertical positions of the center of mass of
the large and small particles respectively. The index varies from 1 (perfectly
segregated: large particles above), to -1 (reversed segregation:
small particles above), and 0 for a homogeneous layer.
 {\color{blue} The evolution of the
instability can be seen on Fig.~\ref{50p20hr15dspatio}(c). For the density 
ratio $\rho_l/\rho=1.5$, the segregation index increases up to 0.6 ($t\simeq 45$s) and subsequently decreases while the system destabilizes to reach 
a stationary value around 0.26. 
The case of a density ratio $\rho_l/\rho=2$ is also reported for comparison. The
destabilization occurs more rapidly but the overall phenomenon is the same.
No instability occurs for the density ratio 1 because
no density gradient counterbalances the upward segregation.  Numerical
simulations for the density ratio 1 have been performed  up  to 400 s, starting
from both initial homogeneous and bi-layer systems. Both systems evolve toward a
stable flat interface and two pure layers of particles, large particles above.}
 {\color{blue} Moreover, experiments in a channel, fed with a mixture of particles with density ratio 1 and size ratios 1.75, 2, 3.5 also show the formation of a uniform stable surface layer composed of large particle and a pure bottom layer of small particles \cite{Thomas00}.
In conclusion, both systems remain axially stable with no banding, which proves the stable state for a density ratio 1, in agreement with a RT instability 
mechanism.}

{\color{blue}
 Our numerical and experimental results are valid in a range of slope angle, while material is flowing (not too low angles), and segregation is happening (high angles lead to a rapid collisional regime). Numerically, the instability has 
been studied from 22$^\circ$ to 26$^\circ$. Larger angle 
leads to an earlier destabilization, but as the flow is more rapid, it 
happens at a longer distance from the start.}

{\color{blue}
The instability have been numerically observed for density ratio as  low as 1.2 for a size ratio 2. 
In granular flows, there is no surface tension which would create a threshold  as in liquids. 
Nevertheless, other stabilizing mechanisms may
occur, as the strong random motion of particles in granular flows. The question of a threshold on the density ratio is still under investigation.}

{\color{blue} Self-induced RT instability could appear paradoxical, but it results from the
competition between two effects, segregation and buoyancy, with
variable intensities. Segregation pushes large particles toward the surface
and buoyancy pushes the dense (and large) particles downward individually, or
collectively. Here, we choose the size and density ratios such that the
segregation is dominant in the mixture.  Consequently, large particles move
upward while surrounded by small ones, and accumulate at the surface where no
small particles are around them. This creates a reverse density
gradient. In the surface layer the segregation effect vanishes because 
large particles are surrounded by large particles.
In this layer, the buoyancy acts in a collective hydrodynamical process
because dense particles are close enough. Consequently, buoyancy dominates
and the whole dense surface layer develops a RT instability.}

{\color{blue} A band pattern also appears in the case of axial segregation in 
partially-filled long rotating drums \cite{Khos00,Finger06,Hill94,caps,Fiedor06,Huang,Santomaso,chen,newey,taberlet,finger15,Choo}. The mechanism, still under
debate \cite{chen}, is likely due to a free surface slope difference between
granular species \cite{Hill94,caps}. But there is no slope angle difference in 
RT instability on inclines. Even though there are some undulations, slope variation
between bands are null in simulations due to periodic boundary conditions and 
could not exceed 1/100$^\circ$ in the channel experiment. Another main 
difference is that in cylinders, axial segregation can happen even with a 
density ratio $\rho_l/\rho=1$ \cite{Khos00,Finger06} contrarily to RT instability on incline
(See Fig.~\ref{50p20hr15dspatio}(c)).}
{\color{blue} Moreover, in drums, large particle bands always form above a core
of small particles \cite{chen,taberlet}. In the RT instability the plumes
made of a mixture and
large particle plumes have an almost symmetrical pattern. The bands
intersect vertically the whole thickness.  The `other' bands 
are composed of a mixture, in opposition to the pure small particle bands 
observed in drums \cite{Hill94,taberlet,finger15}. Finally, the RT instability
is much more rapid to occur. A flow over a distance around 2000$d$ is enough to develop
the instability while in rotating drums, one hundred to several thousands 
rotations are often necessary to
obtain a band pattern \cite{Hill94,Choo,Huang}.}
{\color{blue} The mechanisms leading to the self-induced RT instability on inclines and to axial segregation in tumblers are different.}


\begin{figure}
\center
\includegraphics[width=0.9\linewidth]{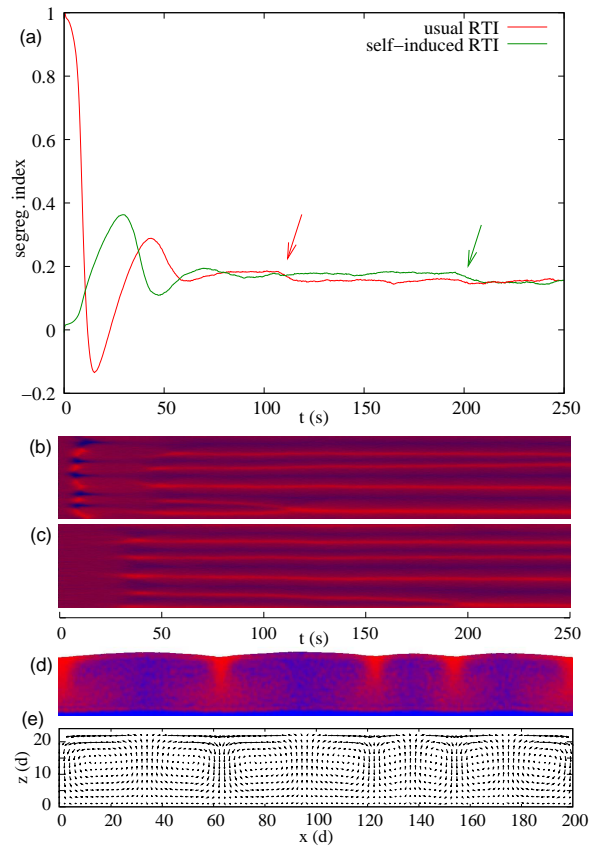}
\caption{Stationary regime. (a) Time evolution of the
segregation index for both initial conditions: two-layers (usual RT instability, red curve)
and homogeneous (self-induced RT instability, green curve). Arrows indicate band merging. 
Space-time diagrams for an initially (b) two-layer flow or (c) one homogeneous-layer flow.
(d) Volume concentration field (large particles in red) 
and (e) corresponding velocity map in the $x-z$ plane of the initially homogeneous flow measured at the end of the simulation.}
\label{50p24compare}
\end{figure}

At least, the long time evolution of the instability is now considered.
The thickness of the granular flows is $H=24d$, the density ratio $\rho_l/\rho=2$ and the length of the flow is reduced to $L=30d$.
Other parameters are unchanged and simulations are conducted for
a longer period, up to $t=250$s.
Figure \ref{50p24compare}(a) compares the 
segregation indexes for the two different initial configurations. 
Both the two-layer system, through a usual Rayleigh-Taylor instability, and
the homogeneous mixture, through a self-induced Rayleigh-Taylor instability, 
converge toward the same steady state for the segregation index
(Fig.~\ref{50p24compare}(a)) and the parallel stripe pattern 
(Figs.~\ref{50p24compare}(b-c)). In these space-time diagrams, bands merging
occurs, indicating that the wavelength in the stationary regime is larger
than those of the initial destabilization.
Each band merging is associated 
to a slight $SI$ decrease (arrows in Fig.~\ref{50p24compare}(a)).
Figures ~\ref{50p24compare}(d-e) show the vertical concentration field
and the corresponding transverse velocity field. 
To obtain a smooth velocity map, averaging over the last 40s of the 
simulation was necessary. {\color{blue} Series of contra-rotative rolls is obtained, corresponding to the position of concentration plumes.}
Figures are extremely analogous to those obtained
in Rayleigh-B\'enard convection cells \cite{Getling}.
It is interesting to note that the maximal transverse velocity is equal to 2\% of the mean velocity in the flowing direction $y$.
Similar behavior with persisting bands and band merging have been observed 
in experiments for low thicknesses.
 {\color{blue} The motor of this convection is the granular segregation 
that drives large particles to the surface even though they are denser. 
As buoyancy drives denser regions downward and segregation upward, the cells are sustained.

Size segregation is then both the cause for the self-induced Rayleigh-Taylor instability
and the self-sustained Rayleigh-B\'enard convection. 
By flowing, the homogeneous layer creates its own unstable state which is quite unusual, but is not able to sustain it and the system evolves to a self-sustained convective state.
{\color{blue} It is interesting 
to note that this very simple system, flowing particles having different
sizes and densities, brings the sufficient mechanisms to induce 
self-organization, pattern formation and instability, features that are
usually met in more complex systems like biological systems \cite{HillPedley05} or complex chemical reactions \cite{Golovin}.
Note also that in a strange way, in this particular configuration, the granular segregation creates an auto-mixing system with large scale heterogeneities, instead of a separating process.}}

We thank S. Gsell and N. Fraysse for interesting comments. 
Centre de Calcul Intensif d'Aix-Marseille University is acknowledged for granting access to its high performance computing resources.

\end{document}